\documentclass[preprint,nofootinbib]{revtex4-1}
\usepackage{amssymb}
\usepackage{amsfonts}
\usepackage{amsmath}
\usepackage{graphicx}%

\providecommand{\U}[1]{\protect\rule{.1in}{.1in}}

\begin{document}
\preprint{ }
\title[Quantifying self-organization]{Quantifying Self-Organization with Optimal Wavelets}
\author{Milo\v s Milovanovi\'{c}}
\affiliation{Mathematical Institute of the Serbian Academy of Sciences and Arts, Belgrade,
11000, Serbia}
\author{Milan Rajkovi\'{c}}
\affiliation{Institute of Nuclear Sciences Vin\'{v}ca, University of Belgrade, P.O. Box 522,
Belgrade,\ 11001, Serbia}

\begin{abstract}
The optimal wavelet basis is used to develop quantitative, experimentally
applicable criteria for self-organization. The choice of the optimal wavelet
is based on the model of self-organization in the wavelet tree. The framework
of the model is founded on the wavelet-domain hidden Markov model and the
optimal wavelet basis criterion for self-organization which assumes inherent
increase in statistical complexity, the information content necessary for
maximally accurate prediction of the system's dynamics. At the same time the
method, presented here for the one-dimensional data of any type, performs
superior denoising and may be easily generalized to higher dimensions.

\end{abstract}
\maketitle






In the most general sense, the term self-organization refers to the process or
processes which cause the emergence of structures and organized behavior
without the external influence. Measuring organization quantitatively has been
the subject of various studies in spite of the inherent difficulties to
characterize complex systems in an accurate manner. The model of
self-organization presented here is inspired by the approach pursued by
Crutchfield and coworkers extending from the early '90s \cite{crutch0},
\cite{cosma2}, \cite{feldman}. Here we adhere to statistical description of
the system and its configurations using the wavelet-domain decomposition and
the properties of the wavelet tree (the graph of wavelet coefficients)
\cite{hernandez}, \cite{mallat} and statistical properties of the wavelet
coefficients. The method is based on a parametric model for a wavelet tree
distribution attributing hidden Markov (HM) variable to each node of the tree.
The wavelet tree is considered as a self-organizing system by identifying
hidden states of wavelet coefficients with local causal states, similar to the
model of self-organization developed in \cite{cosma1} and \cite{cosma2}. Local
complexity in the wavelet-domain is determined as a function of scale and the
global complexity of the tree is utilized as an optimality measure for the
decomposition. Denoising based on the hidden Markov model (HMM) has proven
advantageous over other methods \cite{crouse} and is a natural component of
the method presented here. The method determines the optimal wavelet for
particular data and at the same time evaluates local and global complexity
within the wavelet-based HMM. The method is illustrated using single time
series generated by the dynamic system and it may be easily extended to higher
dimensional data.

The optimality of basis is essential for faithful representation of the
original data (signal) and even more so for compression and denoising. The
only systematic approach to this problem, founded on the microcannonical
cascade formalism and applied to signals with microcannonical cascade
processes, was presented in \cite{oriol} and \cite{oriol2}. Optimal
representation is defined by maximization of mutual information transferred at
successive scales between the wavelet coefficients (parents) at a certain
scale and their descendants (children) at the succeeding one. This method does
not address denoising aspect.

The wavelet transform decomposes a one dimensional spatial signal\footnote{We
chose spatial dependence to avoid possible ambiguity with the notation used
later, but in general time dependence may be used equivalently. }  $f(x)$ in
terms of shifted and dilated versions of a bandpass wavelet function $\psi(x)$
and shifted versions of a lowpass scaling function $\phi(x)$ \cite{hernandez},
\cite{mallat}. For a signal of dyadic dimension $J$ ($2^{J}$ length), the
representation is
\begin{equation}
f=u_{0}\phi_{0}+\sum_{j=0}^{J-1}(\sum_{k=0}^{2^{j}-1}d_{j,k}\psi_{j,k}).
\end{equation}
where $d_{j,k}=<f,\psi_{j,k}>$ and $u_{0}=<f,\phi_{0}>$ while $j$ indexes
dyadic scale of resolution (greater $j$ correspond to higher resolution) and
$k$ indexes the spatial location. For a wavelet $\psi(x)$ centered at
frequency $\xi_{0}$ the detail coefficient $d_{j,k}$ measures the signal
content around place $2^{-j}k$ and frequency $2^{j}\xi_{0}$. Thus, we get a
pyramid of detail coefficients in the form of the binary tree, presented in
Fig. 1(a), in which each coefficient at a resolution scale $j<J-1$ (called
predecessor) has two coefficients at the next resolution scale $j+1$ (called successors)
that share its spatial support. In the following one-index notation for detail coefficients
$d_{j,k}\rightarrow d_{i},i=1\dots I$ is used, starting numeration from the
root of the tree. The label of predecessor for the node $i$ is $\rho(i)$. For random variables we use capital
letters to denote the variable and lower case letters to denote realization of this variable.
Wavelet decomposition of real-world data is sparse so that most of the energy is compacted into small
number of large coefficients, which we call \textit{yang}, while the remaining large number of small
ones we label as \textit{yin}. While yang coefficients provide information on
singularities, yin coefficients carry background information about smooth
characteristics of the data. They also store a significant energy simply
because there are many of them, so their total energy is usually only one
order lower then total energy of yang coefficients. For some deterministic
signals we even observed that yin energy is one order higher than yang energy.
Thus, yin and yang coefficients of a wavelet decomposition are in a kind of
dynamic balance, justifying our choice of terminology.

Sparsity of representation indicates that distribution of wavelet coefficients
is non-Gaussian, typically much more peaky at zero and more spread elsewhere
than a Gaussian \cite{crouse}. A more suitable model of this density is a
mixture of two Gaussians whose components corresponds to yin and yang states:
\begin{equation}
f_{D_{i}}(d)=\sum_{m=1}^{M}P_{S_{i}}(m)g(d,\mu_{i}^{m},\sigma_{i}^{m})
\label{mixture}%
\end{equation}
In the above expression, $f_{D_{i}}$ denotes density function of the random
variable that models detail coefficient of the node $i$, and $P_{S_{i}}$
denotes distribution of hidden variable $S_{i}$ whose values $1$ or $2$
correspond to the yin or yang states of the node. $M=2$ is the number of
components but model can be easily generalized to arbitrary number of hidden
states. Gaussian density function of an argument $d$ with mean $\mu$ and
variance $\sigma^{2}$ is denoted as $g(d,\mu,\sigma)$. An illustration of the
two-state, zero-mean mixture model is presented in Fig. 1(b).

\begin{figure}
[ptb]
\begin{center}
\includegraphics[
width=155mm
]%
{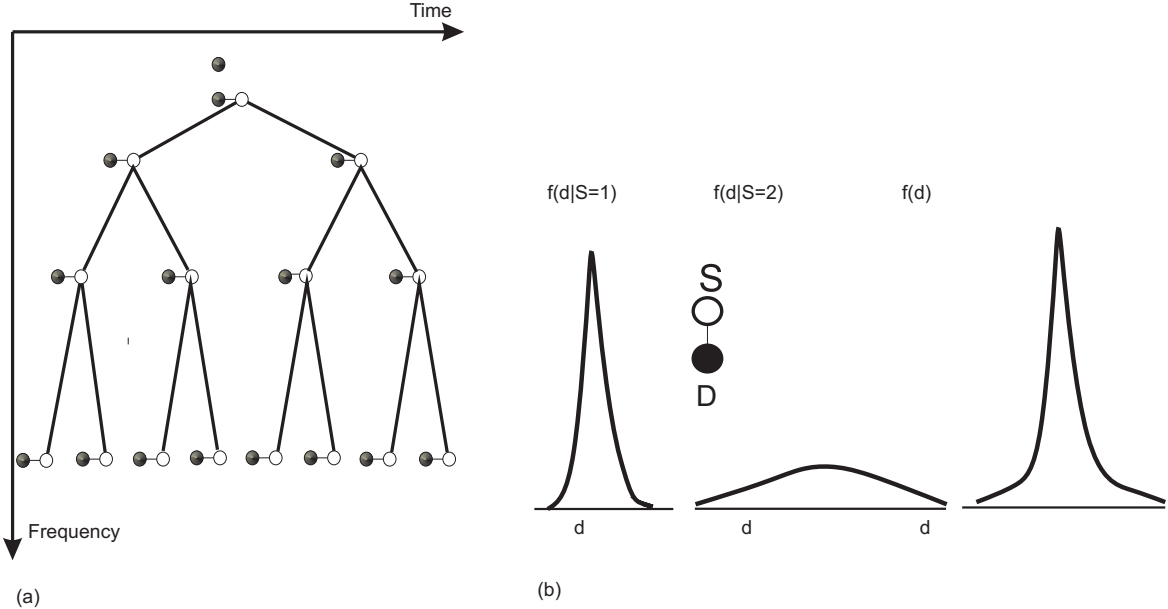}%
\caption{(a)Statistical model of the wavelet transform. Each coefficient $D_{i}$
(black node) is modeled as a mixture with the hidden state variable $S_{i}$
(white node). Hidden states are linked to each other vertically across scales
to yield the Hidden Markov tree (HMT) model.(b)Gaussian two-state mixture model. The model is completely
parametrized by the probability mass function (pmf) of the state variable,
p$_{S}$(1), 1-p$_{S}$(1), and the means and the variances of the two Gaussian
probability density functions (pdf's). The Gaussian conditional pdf's for
\textit{D$\vert$S} are at the left and the center, while the overall
non-Gaussian pdf is on the right.}%
\label{w_tree}%
\end{center}
\end{figure}

Due to the wavelet tree structure, each node at the coarser scale has two
successors at the finer one that share its spatial support. As a consequence,
appearance of yang (yin) coefficient in a node very likely means that its
successors will be yang (yin) coefficients. For that reason, hidden
states tend to propagate across scales (persistence property) \cite{crouse}.
Out of this dependency existing at the hidden state level, detail coefficients
are considered to be decorrelated. Accordingly, dependencies in the wavelet
tree can be completely modeled by conditional probabilities for parent-child
hidden variable pairs. In that way, hidden variables obtain Markov tree
structure which, together with \eqref{mixture}, forms HMM for the wavelet tree
\cite{crouse}.

For $M$-state Gaussian mixture model for each wavelet coefficient
\eqref{mixture}, HMM is determined with parameter model vector
\begin{equation}
{\mathbf{\theta}}=(p_{1}^{m},\epsilon_{i|i\neq1}^{mn},\mu_{i}^{m},\sigma
_{i}^{m}|i=1\dots I;m,n=1\dots M) \label{par}%
\end{equation}
using abbreviations $p_{i}^{m}=P_{S_{i}}(m)$, $\epsilon_{i}^{mn}%
=P_{S_{i}|S_{\rho(i)}=m}(n).$ Parameter estimation is performed by applying
the maximum likelihood principle (ML) which is asymptotically efficient,
unbiased and consistent as the number of observations increases. Direct ML
estimation of the model parameters \eqref{par} from the observed data is
intractable since in estimating ${\mathbf{\theta}}$ we are characterizing the
unobserved (hidden) states ${\mathbf{S}}=(S_{i}|i=1\dots I)$ of the wavelet coefficients
$\mathbf{d}=(d_{i})$. Yet, given the values of the
states, ML estimator of ${\mathbf{\theta}}$ is simple (merely ML estimator of
Gaussian means and variances). Therefore, we employ an iterative expectation
maximization (EM) approach \cite{dempster}, which jointly estimates both the
model parameters ${\mathbf{\theta}}$ and probabilities for the hidden states
${\mathbf{S}}$, given the observed coefficients $\mathbf{d}$.

Due to the limited data available usually from only one or few signal
observations random variables that have similar properties are modeled using a
common distribution or common parameter set, the practice is known as
\textit{tying} \cite{rabiner}. In order to ensure reliable parameter
estimation we must share statistical information between related wavelet
coefficients so we assume that all wavelet coefficients and state variables
within a common scale are identically distributed, including identical
parent-child state transition probabilities. Consequently, in the following
index $j$ in $p_{j}^{m}$, $\epsilon_{j}^{mn}$, $\mu_{j}^{m}$, $\sigma_{j}^{m}%
$ will denote the scale since all parameters of the particular scale are tied
to the same value. The efficiency of the wavelet-domain HMM is demonstrated in
\cite{crouse} by developing a novel signal denoising method. Reconstructing the 
original signal all states with variances less then the noise variance are estimated 
to a single common value i.e. their informational content is completely lost.Having background
noise of unknown power, all yin states of the data are essentially unreliable
and suspected that their content is corrupted by noise. Thus, their content is
certainly preserved only in nearby yang coefficients meaning that optimality
of decomposition implies uniform distribution of yang coefficients in the
wavelet tree.

A paradigmatic approach to the emergence of self-organization phenomena,
presented in \cite{cosma1}, \cite{cosma2} and \cite{feldman} begins with a
dynamic random field on the network on which the random field of local
causal states is constructed. To predict the original field either
locally or globally, it is sufficient to know causal states. We find that this
model shares common features with the wavelet-domain HMM and extend this
analogy to a new level. The starting point in analyzing and predicting
observations is to regard them as distorted measurements of another, unseen
set of state variables which have their own dynamics. We comply with the
framework of \cite{grassberger}, where the complexity is the minimal amount of
information about the system's state needed for optimal prediction and
further follow the idea of \cite{crutch0} to identify the complexity of a
system with an amount of information needed to specify its causal state, the
quantity labeled as statistical complexity. Following \cite{cosma1} and
\cite{grassberger} the local statistical complexity is defined as the entropy
of local causal state
\begin{equation}
C(x,t)=H(S(x,t)).
\end{equation}
If a spatially stationary process is dynamically autonomous from external
influences self-organization takes place between time $t$ and time $t+T$ if
and only if $C(x,t)<C(x,t+T)$ \cite{cosma1}. Our aim is to perceive HMM from
the viewpoint of self-organization giving the concept of self-organization
specific physical interpretation within the model. Some semantic analogies of
the terms used in \cite{cosma1} and \cite{cosma2} and the wavelet-domain HMM
will be used in order to make the ideas more clear. First, it is necessary to
define the time axis. Interdependence of the nodes takes place vertically
through the tree (persistence property)\ so we consider time axis as dyadic
frequency axis directed from the coarsest to the finest scale. We regard
signal domain as spatial even for temporal signals because the concept of time
is replacing the frequency domain. Thus, by introducing \textit{diffeomorphism
invariance }the wavelet tree becomes the spatio-temporal tree. The direction
of time is determined by the branching process representing information flow
from parent to descendant coefficients. In the context of binary tree
structure and the chosen time axis causality is defined by interdependence of
the wavelet coefficients so it lies solely in the HM structure of the wavelet
tree. Tying in the EM algorithm implies stationarity (and vice versa) in the
spatial domain. Due to persistence property causality, considered as an
optimal prediction of the wavelet tree containing information about yin and
young states, is defined by presence or absence of singularity in the spatial
support of wavelet coefficients. Therefore, hidden state variables $S_{i}$ are
considered as local causal states which form the \textit{wavelet machine} or
\textit{w-machine} in analogy with the $\epsilon-$ \textit{machine}%
\footnote{Note that the \textit{w-machine} does not satisfy the unifilarity
property of $\epsilon-$ \textit{machines.}} presented in \cite{crutch1} and
\cite{crutch2}. Random variable $\mathbf{S}=(S_{i})$ represents
the global causal state which contains minimal information for optimal
prediction in the spatial domain. The proof follows from the EM algorithm which
minimizes $H(\mathbf{S}|\mathbf{d})$ so we have $\mathbf{S}=f_{\theta}(\mathbf{D})$. Knowledge
of $\mathbf{S}$ is related to optimal prediction because $\mathbf{D}$ in HMM depends on $\mathbf{S}$
only. The entropy of the wavelet tree may be expressed as
\begin{equation}
H(\mathbf{D})=H(\mathbf{D},\mathbf{S})=H(\mathbf{D}\left\vert \mathbf{S}\right)  +H(\mathbf{S}),\label{ent}%
\end{equation}
where $H(\bf{D})$ and $H(\bf{D}\left\vert\bf{ S}\right)  $ are differential entropies of
continuous random variables. The extensive term $H(\bf{D}\left\vert \bf{S}\right)  $
represents irreducible randomness that remains even after all correlations are
subsumed. Addition of noise increases only this term while complexity $H(\mathbf{S})$
remains unaltered. Local complexity $C_{i}=H(S_{i})$ has a specific physical
interpretation - it is higher if the distribution of hidden yang an yin states
in the node is more uniform. In that case, there is higher probability of yang
coefficient appearance based on the persistence property in the nodes at the
immediate neighboring scales meaning that information stored in $D_{i}$ will
be preserved. Yet, it should be noted that local causal state in this model is
statistic of the whole tree $\mathbf{D}$, thus separation into future and past
becomes irrelevant for causality. Local causality implies both prediction and
retrodiction and this property of the model we call \textit{temporal
irrelevance}. We indicated that local complexity $C_{i}=H(S_{i})$ is the
measure which guarantees that the information contained in the node is
optimally preserved. Global complexity $C=H({\mathbf{S}})$
fulfills that goal for the complete tree. Higher global complexity means that
yang states are more uniformly distributed within the tree allowing for more
optimal preservation of background information. So, we define optimal
representation of the data (signal) as the one which maximizes global
complexity of the tree. We note that factorization of global causal state into
local ones in the wavelet HMM is different from the model presented in
\cite{cosma1} because global state is not determined from local states in only
one time instant. This is the consequence of temporal irrelevance since
prediction takes into consideration the complete signal, i.e. both the past
and the future of the wavelet tree. Regardless of these differences, we
demonstrate that optimality of decomposition is related to the increase of
local complexity and thus to the self-organization.

Derivation of the global complexity in terms of model parameters yields
\begin{equation}
C=H({\mathbf{S}})=\sum_{m}-p_{0}^{m}(logp_{0}^{m}+\sum_{n}2\epsilon_{1}%
^{mn}(log\epsilon_{1}^{mn}+\sum_{r}2\epsilon_{2}^{nr}(log\epsilon_{2}%
^{nr}+\dots)))
\end{equation}
This expression takes higher values if conditional variables $S_{i}%
|S_{\rho(i)}=m$ are more uniformly distributed i.e. if probability of changing
state is higher. But in this case local states also tend to be more uniformly
distributed so that local complexity increases. It is also related to
successful denoising using algorithm presented earlier, because higher
complexity suggests more uniform distribution of yang coefficients and so
information contained in the yin coefficients, which are more affected by
noise, is preserved better. We have tested the model on a variety of signals
and here we include the \textit{y}-component of the Lorentz chaotic
oscillator. White Gaussian noise of variance equal to 1 is added to the
signal. The energy density of the remaining noise is estimated after
denoising. Increase of local complexity in temporal domain is evaluated as
maximal length of the interval at which the complexity function increases
monotonically. In Table 1 we present results for the \textit{y-}component of
the Lorentz chaotic oscillator. The entropy is normalized so that it is
bounded between 0 and 1. Representatives from the standard wavelet families
are included, namely Haar (haar), Daubechies (db2), Symlet (sym3),
Coiflet(coif1), Biorthogonal (bior1.3), Reverse Biorthogonal (rbior1.3) and
Discrete Meyer (dmey). Biorthogonal wavelets are named as Biorn1.n2 where n1
is the number of the order of the wavelet or the scaling function and n2 is
the order of the functions used for decomposition. Brief inspection of Table 1
suggests the discrete Meyer wavelet (dmey), marked in bold, as the optimal choice. It should be
emphasized that energy density of the remaining noise is not an indicator of
optimality of representation, because optimal representation is a general
concept independent of particular signal processing application. However, it
is obvious that optimality of representation based on self-organization in the
wavelet-tree implies optimal wavelet-based noise reduction.
\begin{center}
\begin{table}[th]%
\begin{tabular}
[c]{|c|c|c|c|c|c|c|c|}\hline
wavelet & haar & db2 & sym3 & coif1 & bior1.3 & rbio1.3 & dmey\\\hline
\multicolumn{1}{|l|}{remaining noise} & 0.6138 & 0.3888 & 0.3234 & 0.3821 &
0.6442 & 0.3142 & \textbf{0.2559}\\
\multicolumn{1}{|l|}{global complexity} & 0.2984 & 0.6474 & 0.7300 & 0.6507 &
0.2350 & 0.6795 & \textbf{0.8075}\\\hline
\end{tabular}
\par
$\quad$\newline Table 1.\end{table}
\end{center}

We illustrate the method in the context of dynamical systems by considering structure
and randomness of the time series generated by the logistic map on the unit interval
$f(x)=rx(1-x)$, where $r\in\lbrack0,4]$. The term $H(\mathbf{S})$ in Eq.\eqref{ent}
represents the measure of complexity (structure) and the conditional entropy
$H(\bf{D}\left\vert\bf{ S}\right)  $ is the measure of randomness. Both are represented
in Fig. 2 as a function of parameter $r$ generated using the
optimal, biorthogonal1.3, wavelet. The maximum complexity is attained for
parameter value 3.5926, i.e. the value at which the deterministic chaos sets
in. In Fig. 3 we present the complexity-entropy diagram
corresponding to the $r\in\lbrack2.8,4]$ parameter region.
\begin{figure}
[ptb]
\begin{center}
\includegraphics[
height=6.2955cm,
width=12.2681cm
]%
{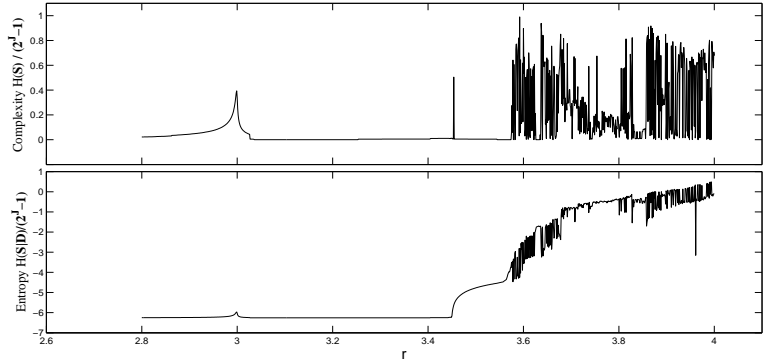}%
\caption{Complexity $H(S)/(2^{J}-1)$ and entropy rate $H(D|S)/(2^{J}-1)$ as a
function of the parameter r. The r values were sampled uniformly in increments
of 0.0001. Note the negative entropy values as a consequence of the differential entropy property.}%
\label{Fig2}%
\end{center}
\end{figure}
\begin{figure}
[ptb]
\begin{center}
\includegraphics[
height=5.731cm,
width=11.1632cm
]%
{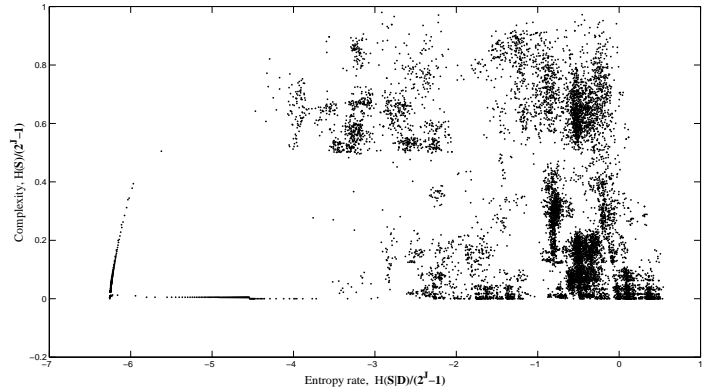}%
\caption{Entropy rate and complexity pairs $(H(D|S)/(2^{J}-1),H(S)/(2^{J}-1))$
for the logisitic map. The parameter $r$ values were sampled uniformly in
increments of $0.0001$. Negative entropy values stem from the properties of
differential entropy that takes all values from $\mathbb{R.}$}%
\label{Fig3}%
\end{center}
\end{figure}

For a given value of entropy multiple values of complexity are noticed indicating an intricate
relationship between these two quantities. Not all complexity values are
realizable for a particular entropy rate. Organization is evident in the
diagram consisting of low and very high density regions exhibiting
self-similar structure in the central part of the diagram. Both the lower and
the upper bounds are well defined.

We have argued that \textit{w-machine} establishes relationship between
information, prediction, retrodiction and denoising founded on the choice of
the optimal wavelet and within the framework of statistical mechanics.
Statistical complexity may be reliably calculated from data and at the same
time noise may be removed in a highly efficient manner. The method can be
easily adapted to 2-dimensional signals.

The authors acknowledge support by the Serbian Ministry of Education and
Science through the projects OI 174014 and III 44006.

\end{document}